\documentclass[12pt]{article}

\newcommand{\be}{\begin{equation}}
\newcommand{\ee}{\end{equation}}
\newcommand{\bea}{\begin{eqnarray}}
\newcommand{\eea}{\end{eqnarray}}
\newcommand{\br}{\hskip .25cm/\hskip -.25cm}

\newcommand{\nonu}{\nonumber\\}

\newcommand{\ol}{\overline}
\usepackage{graphicx}

\begin{document}

\begin{titlepage} 

\begin{center}
{\Large{\bf Anomalous magnetic moment in parity-conserving $QED_3$}}
\end{center}

\vspace{0.2in} 
\begin{centering} 

{\bf Jean Alexandre}$^a$, {\bf K. Farakos$^b$} and {\bf N.E.~Mavromatos}$^a$

\vspace{0.2in}
$^a$ King's College London, Department of Physics, Theoretical Physics, Strand WC2R 2LS, U.K. 

$^b$ National Technical University of Athens, Zografou Campus, Athens 157 80, Greece.

\vspace{0.2in}
{\bf Abstract} 

\end{centering} 

In this article we derive the anomalous magnetic moment
of fermions in (2+1)-dimensional parity-conserving QED$_3$,
in the presence of an externally applied constant magnetic field.
We use a spectral representation of the photon propagator to 
avoid infrared divergences. 
We also discuss the scaling with the magnetic field intensity
in the case of strong external fields, where 
there is dynamical mass generation  for fermions
induced by the magnetic field itself (magnetic catalysis). 
The results of this paper may be of relevance to the physics of 
high-temperature superconductors.  

\end{titlepage} 

\section{Introduction}

Three dimensional gauge theories find interesting 
physical applications
to the physics of planar high-temperature superconductors 
and more general doped planar antiferromagnets~\cite{mavpap}. 
On the other hand, three dimensional field theories
may be used as toy models for four dimensional physics in many respects, 
for instance to get some information on 
the treatment of non perturbative effects of gauge 
theories~\cite{kovner}.

Odd dimensional field theories are characterised 
in general by the anomalous breaking of the discrete symmetries of Parity (P) 
and/or Time reversal (T), and this may have profound physical 
consequences for the relevant systems. For instance there is a long
and, in our opinion, 
still outstanding debate as to whether parity is broken spontaneously
or anomalously by the ground state of high-temperature superconductors,
whose physics may be modeled in some respect by three-dimensional 
gauge field theories. 

In this last respect we mention that current theoretical 
models of high-temperature 
superconductors are based on the fact that such systems are characterised
by nodal points in their surfaces, i.e. points in momentum space where 
some energy gap function  vanishes. Linearisation about such points
results in relativistic models of high temperature superconductivity,
which have been proposed some time ago~\cite{dorey}, 
and recently have been revived 
from a rather different perspective~\cite{herbut}.

An important feature of the antiferromagnetic nature of the 
system, or the way the continuum field-theoretic limit 
is taken, is the doubling of the relevant fermion species
which represent charged degrees of freedom in a spin-charge
separation framework. Indeed, the continuum degrees of freedom 
of the resulting (2+1)-dimensional field theory are 
two 2-component fermions, 
$\Psi_1,\Psi_2$, coupled to a statistical gauge field, representing
magnetic interactions in the underlying condensed-matter system, 
and interacting with external electromagnetic field, since the fermions
(holons) represent charged excitations~\footnote{In some other 
approaches~\cite{lee}
these fermions are electrically neutral, and it is the bosons in a spin-charge 
separation framework that are electrically charged. In the present article
we shall not discuss such models.}. 

The presence of 
an even number of fermion species 
implies a parity-conserving model as far as the fermion mass is concerned.
This is because in that case, parity can be defined in an extended way 
so as to exchange fermion species, thereby leading to a parity-invariant mass.
The two component fermions may be combined to a four-component spinor 
$\psi = \left(\Psi_1, \Psi_2\right)$, 
in which case a parity invariant mass term 
is simply $m {\overline \psi} \psi$. Depending on the details of the 
underlying statistical model one may also have more than one ``flavour'' 
of these four component spinors. 

The dynamical generation of a parity -invariant mass for fermions seems to be 
preferred energetically due to the Vafa-Witten theorem~\cite{vafa}, 
in the absence of 
externally applied fields. This defines then a P-invariant ground state.
However, from a physical point of view high-temperature superconducting 
materials are strongly type II superconductors, which implies that 
when an external magnetic field is applied, the magnetic lines will
penetrate the material in its superconducting phase significantly. 
As a consequence, it makes sense to consider the effects of an 
external magnetic field on holons in their superconducting state. 
This brings up the question as to whether a parity-violating mass 
can be generated dynamically under the influence of the external
field, since in that case the Vafa-Witten theorem is inapplicable. 

Experimentally there are indications that parity violation (in the form of
resultant edge chiral currents in the superconducting planar materials) 
may occur in such circumstances. 
According to the models of \cite{dorey,fm}, however, a superconducting 
phase is characterised
by massive charged fermions (holons) with a parity-conserving mass.
In fact it has been argued  
that it is the nodal relativistic fermions that play a crucial role 
in the passage from the pseudogap to the superconducting phase~\cite{asm}. 
Even in the presence of a magnetic field, which catalyzes the fermion
mass generation, the parity conserving mature of the magnetically-induced 
fermion mass gap can be maintained in a consistent way~\cite{fkm}. 

It is therefore natural to inquire whether there is a possibility 
that parity is violated upon the application of external magnetic fields
in the effective field theoretic continuum action
of high temperature superconductors, as the experiments seem to indicate, 
but this violation 
is independent of the superconducting mass gap.
It has been proposed in \cite{momen} 
that the answer to this question may come from the magnetically-induced
anomalous magnetic moment of the fermions (holons), which from an effective
action point of view contributes a non-minimal term that violates parity and time reversal, 
as a result of the fixed direction of the external 
magnetic field. In \cite{am} a more elaborate situation 
of an induced magnetic moment
for fermions has been calculated, in a case in which the 
gauge field is considered four dimensional but strongly anisotropic. 
These calculations showed that there is indeed an induced magnetic moment
for the fermions, which violates parity and time reversal, 
scaling with the externally applied field in a way that can be tested 
experimentally.  
However, those results were not complete 
in the sense that the definition for 
the magnetic moment used was the one that is 
usually applied to the weak field case. 

In this article we complete the computation by considering the 
appropriate definition for the magnetic moment 
in both strong and weak fields.
We consider only the 
case in which the field theory lives exclusively on a plane, and compute
the corresponding anomalous magnetic moment for fermions in 
the case of magnetic fields applied perpendicularly
to the plane where the fermions live. 
This computation has a field theoretic interest in its own right, 
independently
of the possibility of applying the result to the physics of
condensed-matter systems such as antiferromagnets and superconductors. 
As we shall see, the issue of the removal of infrared (IR) divergences
by means of spectral decomposition of propagators is non trivial in this case. 
For these reasons we have written the article in a language
suitable for field theorists, reserving a possible connection with 
condensed matter physics only for the conclusions. We intend to 
come back to a detailed condensed-matter analysis in a future publication.

The structure of the article is as follows: Section 2 
presents the computation of the anomalous magnetic moment $\mu_B$
in the usual case, i.e. in the limit
of vanishing external field. This computation follows the well-known derivation in 3+1 dimensions, but needs more 
care in order to deal with a logarithmic IR divergence. This is done by using a spectral representation of the 
dressed photon propagator. This method, originally set up in \cite{jackiw}, cures the would-be IR divergence by taking
into account a partial resummation of higher order quantum corrections and leads to a non-analytic dependence 
in the coupling constant. We discuss two separate cases: (i) weakly coupled
gauge theories with a large (compared to the coupling) bare fermion mass,
and (ii) strong coupling gauge models, with a small (compared to the coupling)
dynamically generated mass of the fermions, in a large-$N_f$ treatment,
with $N_f$ the number of flavours of four-component spinors.
We also compare briefly our results with some relevant results 
on induced magnetic moments in 
parity violating gauge models of anyons. 
Section 3 deals with the computation of the anomalous magnetic moment in the presence of 
a strong external 
magnetic field. In this case we have to come back to the general definition of $\mu_B$ using the fermion self energy
computed with the non-perturbative fermion propagator in the presence of an external magnetic field. 
We will show that the anomalous magnetic moment can be defined in the lowest Landau level, provided we are 
in the rest frame of the massive fermion. $\mu_B$ also exhibits a logarithmic divergence that is cured with 
the same spectral representation method as in Section 2.
We also consider in this case the scaling of the anomalous magnetic moment with the strong magnetic field. For this
we consider, for the fermion mass, the dynamical mass $m_{dyn}$ that is generated by the external field, using 
previous works \cite{fkm,afk} where the dependence of $m_{dyn}$ on the magnetic field was set up. Conclusions and 
a possible connection of 
our results with the physics of high temperature superconductors
are presented in Section 4.

\section{Magnetic moment in weak field}

\begin{figure}[t]
\centering
\includegraphics[width=8cm]{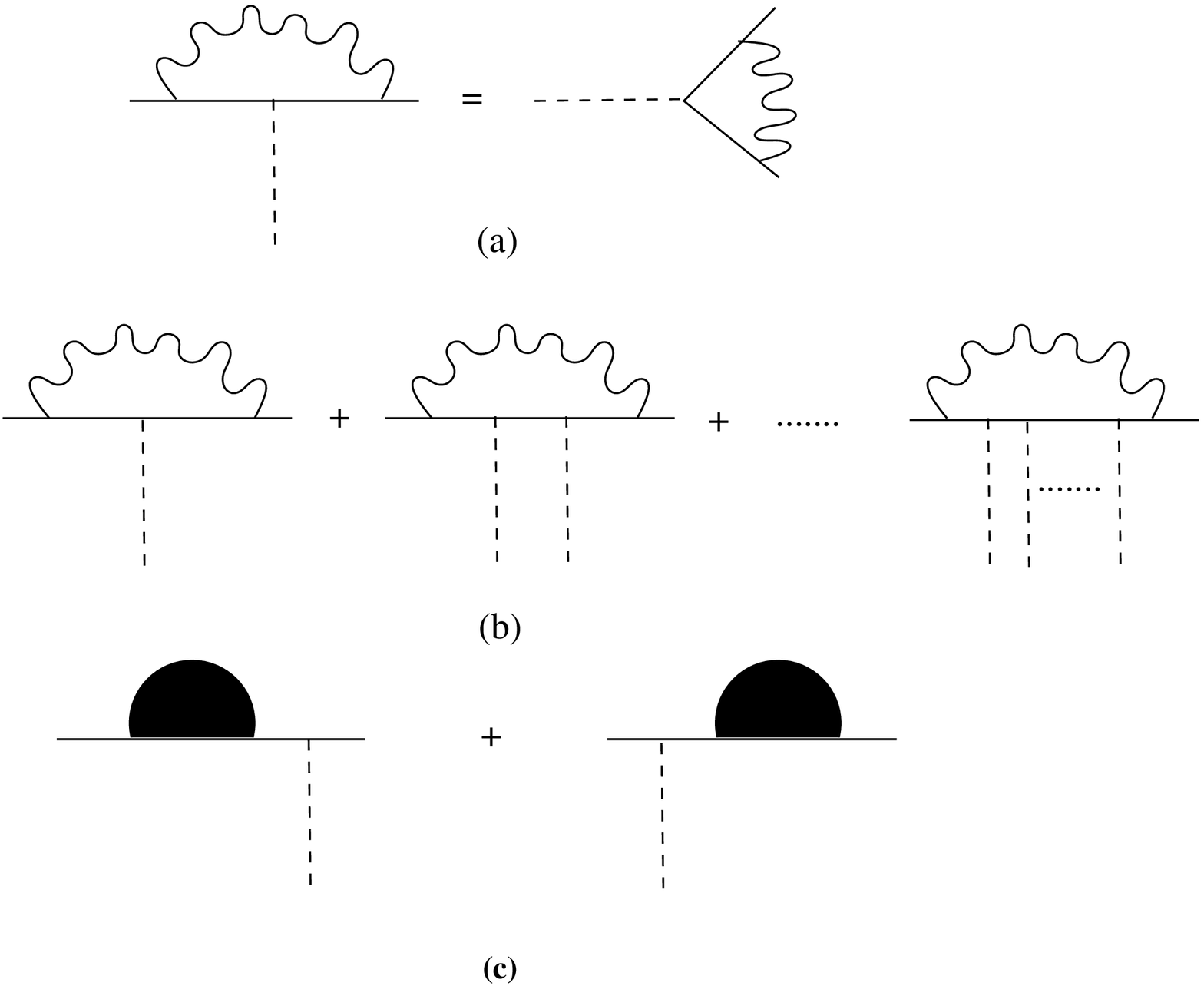}
\caption{{\it One-loop graphs used for the computation of the anomalous 
magnetic moment. The fermions are represented 
by a continuous line, the dynamical gauge boson 
by a wavy line and the external field by a dashed line}.
(a) {\it In the case of a weak external magnetic field,
the fermion self energy corresponds to the vertex correction since there is only one external 
photon insertion.} (b) {\it In the case of a strong magnetic field, the fermion self energy corresponds to
an infinite series of graphs, where one sums over 
the external field propagators. This summation is taken into
account by the Schwinger proper-time representation of the fermion propagator in a classical electromagnetic field.}
(c) {\it Self-energy corrections to the external fermion polarization
spinors. Such reducible graphs should be added to (a) or (b), but are
proportional to $\gamma^\mu$, and hence do not contribute to the anomalous magnetic moment. They provide finite renormalization of the vertex, 
canceling the terms proportional to $\gamma^\mu$ in (\ref{Gamma}).
The dark-filled semicircles indicate dynamical gauge boson corrections with ((b)) 
or without ((a)) external 
field lines.}}  
\label{fig:vertex}
\end{figure} 
 
In the case of a weak external magnetic field $\vec B$, the magnetic moment $\mu_B$ is defined as 
the coefficient of the first
order, linear, interaction between $\vec B$ and the spin of the fermion \cite{weinberg}. 
$\mu_B$ can be found by computing the quantum corrections to the vertex 
$\Gamma^\rho$ representing the interaction of $A_\mu^{ext}$
with the fermion $\psi$, which is taken as a 4-component fermion including the
two flavours $\Psi_1$ and $\Psi_2$ 
(see fig.~\ref{fig:vertex}(a)~\footnote{It is understood that 
one should add to these (one-particle irreducible) 
graphs also one-particle reducible
graphs (c.g. fig.~\ref{fig:vertex}(c)), where the 
external fermion lines get self energy corrections, expressing the 
effects of wave function renormalization to the 
external fermion polarisation tensors.
Such corrections are proportional to $\gamma^\mu$ and do not contribute to
the anomalous magnetic moment part of the vertex (\ref{Gamma}), 
which is linear in momenta 
$(p + q)^\mu $. This will always be understood in what follows, and it is valid
for all ranges (strong, intermediate and weak)
of {\it both } the gauge coupling and the external field.
To be complete in our statements, therefore, we say that   
the sum of these graphs and those of fig.~\ref{fig:vertex}
yields formally the physically observable 
contributions to the anomalous magnetic moment.}). 
In the limit 
where the momentum flowing from the external field vanishes,
the on-shell vertex reads: 
\be\label{Gamma}
\Gamma^\rho_{on-shell}(p,q)={\cal Z}\gamma^\rho-\frac{\mu_B}{2m}(p^\rho+q^\rho),
\ee
where ${\cal Z}$ is the quantum correction to the minimal coupling, $m$ 
is the fermion mass, and $g$ is the
coupling to the dynamical gauge field~\footnote{Throughout this work we keep 
$g$ different from the real (three-dimensional) electromagnetic coupling 
$e$. This allows for a straightforward extension of the present analysis 
to models of relevance
to high temperature superconductors~\cite{dorey,fm,herbut}, 
where the statistical photon,
distinct from the real electromagnetic one, expresses
effectively magnetic interactions among the relevant degrees of freedom.}.
In this expression, $p,q$ are the momenta of the incoming and outgoing fermion, 
i.e. we consider the limit $p\to q$. Using Dirac equation, the on-shell vertex (\ref{Gamma}) leads 
then to the expected anomalous magnetic moment term in the effective lagrangian
\be\label{mub}
\frac{\mu_B}{2m}\ol\psi\sigma^{\rho\kappa}\psi~\partial_\rho A^{ext}_\kappa,
\ee
which corresponds to a non-minimal parity-violating coupling generated by quantum effects.

Before proceeding with the relevant analysis we consider at this stage
useful to point out that in our model the limiting 
relativistic 
velocity $v_F$ of the fermions and dynamical gauge bosons 
is not, from a physical viewpoint, 
the real speed of light $c$, but rather the 
velocity of the node of the fermi surface of an underlying 
microscopic condensed matter model~\cite{dorey,fm,herbut}
whose low energy effective theory near the node 
is described by the ``relativistic''
gauge model under consideration. For realistic 
systems~\cite{dorey,fm,herbut,lee} one has (at most) $v_F \sim 10^{-4}c$,
which means that in our system of units $v_F=1$ we have 
$c \sim  10^4$ (at least).
This implies that~\cite{fm} in this model the real electron charge 
${\tilde e}$ 
is related to the effective three-dimensional charge $e$ appearing in the model
by 
\begin{equation}
e = {\tilde e}/c~, 
\label{rescale2}
\end{equation}
which in turn means that 
the physically observable value of the anomalous magnetic moment
should be given by rescaling the value $\mu_B$ in (\ref{mub}) by $1/c$:
\begin{equation}
\mu_B^{\rm phys} = \frac{1}{c}\mu_B
\label{rescaled}
\end{equation}
This will always be understood throughout this work. 

In what follows we shall distinguish two cases: (i) a bare fermion mass, and 
weak gauge coupling $g^2 \ll m$, (ii) a sufficiently strong gauge coupling
$g^2$, responsible for the dynamical generation of fermion mass
$m \ll g^2$. In case (ii) the mass scales with the weak magnetic field $B$,
as shown in \cite{fkmm}, as a result of the well-known 
magnetic catalysis phenomenon~\cite{gusynin}.

\subsection{Weak Gauge Coupling and Bare Fermion Mass}

We consider the Feynman gauge and assign the mass $M$ to the dynamical vector field to 
regularize the integrals involving an infrared divergence. The latter will then be avoided using a
spectral representation of the photon propagator.
 
The one-loop correction to the vertex (c.f. fig.~\ref{fig:vertex}(a))
is, in the limit of zero incoming momentum from the external field,
\be\label{init}
\Gamma^\rho (p)=-ig^2\int\frac{d^3k}{(2\pi)^3}\frac{\gamma^\mu(\br p+\br k+m)
\gamma^\rho(\br p+\br k+m)\gamma_\mu}
{[(p+k)^2-m^2]^2[k^2-M^2]}.
\ee
For on-shell fermions, we have $p^2=m^2$, such that
\bea
&&\Gamma^\rho (p)=-\frac{2ig^2}{(2\pi)^3}\int_0^1 dx\int_0^{1-x}dy\nonu
&&\times\int d^3k
\frac{\gamma^\mu(\br p+\br k+m)\gamma^\rho(\br p+\br k+m)\gamma_\mu}{\{[k+(x+y)p]^2-(x+y)^2m^2-(1-x-y)M^2\}^3}.
\eea
The vertex appears as $\ol\psi\br\Gamma\psi$ in the Lagrangian, such that the momentum $\br p$, when appearing on the
left or the right of the group of matrices $\gamma^\mu,\gamma^\nu,\gamma^\rho$ can be replaced by $m$ since 
for on-shell fermions we have $\br p\psi=m\psi$. One obtains then for the terms proportional to $p^\rho$
\be
\frac{4\pi i\times 2ig^2}{(2\pi)^3}\int_0^1 dx\int_x^1 dz \int_0^\infty dq_E~q_E^2
\frac{8z(1-z)mp^\rho}{[q_E^2+z^2m^2+(1-z)M^2]^3},
\ee
where $q_E=k_E+zp_E$ is the Euclidean momentum and $z=x+y$. The integration over $q_E$ gives finally
the following expression for the magnetic moment
\be\label{massphot1}
\mu_B^{(0)}(m,M)=\frac{g^2 m^2}{2\pi}\int_0^1 dx\int_x^1 dz\frac{z(1-z)}{[z^2m^2+(1-z)M^2]^{3/2}}.
\ee
Here the subscript $(0)$ stands for the naive magnetic moment, where the infrared divergence at $M=0$ has 
not yet been taken into account. 
We will come back to this divergence after integrating over $z$ and $x$ for finite $M$.

First, we can see that in the case $m=0$ and $M\ne 0$, the integration over $x,z$
is convergent and the anomalous magnetic moment vanishes:
\be
\mu_B^{(0)}(0,M)=0.
\ee
But we are more interested by the case $m\ne 0$ and $M=0$. For this we first keep $M\ne 0$. 
We obtain then: 
$\mu_B^{(0)}=(g^2/2\pi m)(J_1+J_2)$ where, in the limit $\beta\to 0$, $J_1$ has a finite value
and $J_2$ contains an infrared logarithmic divergence. We have, for any non-vanishing 
value of the parameter $\beta=M^2/m^2$,
\bea\label{j12}
 && J_1 = \int_0^1 dx\left\{   
\frac{3-\beta}{2-\beta/2}\left(1-\frac{\beta}{2}-
\frac{x-\beta/2}{\sqrt{x^2+(1-x)\beta}}\right) + \right.\nonumber \\
&& \left. \beta-1- \ln\left(\frac{4-\beta}{2x-\beta+2\sqrt{x^2+(1-x)\beta}}
\right)\right\} = \nonumber \\ 
&& \frac{3\beta + 3\sqrt{\beta} -4}{\sqrt{\beta} + 2} 
- \frac{\beta}{2}{\rm ln}\left( 1 + \frac{2}{\sqrt{\beta}}\right)
\nonumber \\ 
&& J_2 =\int_0^1 dx\frac{1-\beta}{\sqrt{x^2+(1-x)\beta}}
=(1-\beta)\ln\left(1+\frac{2}{\sqrt\beta}\right).
\eea 

To avoid the divergence of $J_2$ in the limit $M\to 0$ we proceed as in \cite{jackiw}, where the authors
compute higher-loop corrections to the photon and fermion propagators in massless $QED_3$.
Instead of the bare
photon propagator that was used in Eq.(\ref{init}), we consider the following spectral representation
of the dressed propagator for massless gauge boson
\be\label{photprop}
D_{\mu\nu}(k)=g_{\mu\nu}\int_0^\infty dM\frac{\rho(M)}{k^2-M^2},
\ee
where the weight $\rho(M)$ includes the quantum corrections. It is known that the one-loop 
vacuum polarization is \cite{appelquist}
\be\label{poltens}
\Pi_{\mu\nu}(k)=\frac{g^2}{4\pi}\left(g_{\mu\nu}-\frac{k_\mu k_\nu}{k^2}\right)
\left\{2m+\frac{k^2-4m^2}{k}\sin^{-1}\left(\frac{k}{\sqrt{4m^2+k^2}}\right)\right\},
\ee
where $k=\sqrt{k^2}$, such that, in the limit of small momentum $k$ 
\be\label{dressphotpropapprox}
D_{\mu\nu}(k)=\frac{g_{\mu\nu}}{k^2-\frac{g^2}{6\pi}k}
\ee
Note that there is no branch cut problem in the complex plane $k$: the propagator (\ref{dressphotpropapprox})
is an approximate one and the only pole that occurs when taking the full polarization tensor (\ref{poltens})
into account is for $k=0$. Note also the non-perturbative nature of the infrared quantum corrections since 
$g^2k>>k^2$ when $k\to 0$.

To take into account the correction (\ref{dressphotpropapprox}) for small momenta, 
which is sufficient to cure the IR divergence, the function $\rho(M)$ should be taken as follows:
\be\label{rho}
\rho(M)=\frac{\frac{g^2}{3\pi^2}}{M^2+\left(\frac{g^2}{6\pi}\right)^2}.
\ee

Taking higher order correction with this method cures the infrared divergence of the magnetic moment since the 
former is logarithmic and thus integrable. The expression for the anomalous magnetic moment for massless
gauge boson is thus given by
\bea\label{mubfinal}
\mu_B(m,0)&=&\int_0^\infty dM\rho(M)\mu_B^{(0)}(m,M)\\
&=&\frac{g^2}{2\pi m}\int_0^\infty dM\rho(M)\left\{J_1(m,M)+J_2(m,M)\right\}.\nonumber
\eea
Since the weight $\rho$ is normalized, i.e. $\int_0^\infty dM\rho(M)=1$, we have
\bea
\mu_B(m,0)&=&\frac{g^2}{2\pi m}\left\{J_1(m,0)+\int_0^\infty dM\rho(M)J_2(m,M)\right\}+{\cal O}(g^2/m)^2,\nonu
&=&-\frac{g^2}{\pi m}+\frac{g^2}{2\pi m}\int_0^\infty dM\rho(M)J_2(m,M)+{\cal O}(g^2/m)^2,
\eea
where ${\cal O}(g^2/m)^2$ are higher order terms in $g^2/m$.
An expansion in powers of $g^2/m$ leads then to
\be
\int_0^\infty dM\rho(M)J_2(m,M)=-\ln\left(\frac{g^2}{12\pi m}\right)+{\cal O}(g^2/m),
\ee
where we used the identity 
\begin{equation}\label{identity}
\int_0^\infty dy\frac{\ln y}{1+y^2}=0~.
\end{equation}

Finally, the divergence-free anomalous magnetic moment is, in the case of massless dynamical gauge field
and massive matter field $m\ne 0$
\be\label{muB}
\mu_B(m,0)=-\frac{g^2}{\pi m}+\frac{g^2}{2\pi m}\ln\left(\frac{12\pi m}{g^2}\right)+{\cal O}(g^2/m)^2.
\ee
Note that, as expected from this spectral representation method, the would-be infrared divergence is cured 
by the non-analytical dependence $\varepsilon\ln\varepsilon$ in the dimensionless coupling $\varepsilon=g^2/m$. It is understood that the 
rescaling (\ref{rescaled}) is in operation if we wish to compute the physically
measurable value.

\subsection{Strong Gauge Coupling and Dynamically Generated Fermion Mass}

In the remainder of this section we would like to discuss briefly
the case where $g^2 \gg m$, which is the case where one 
has a dynamically generated mass. This case is also 
of great physical interest due to its direct application to
high temperature superconducting models~\cite{dorey,fm}.
A complete treatment of this case is not yet available, nevertheless
one can get a consistent and satisfactory for our purposes 
treatment within the so-called large 
$N$-approximation~\cite{appelquist,largen,nash}.

To this end, we assume there are $N_f$ four-component fermionic flavours,
and one fixes the dimensionful gauge coupling of the statistical gauge field
(not the real electromagnetic one) as: $\alpha = N_f g^2/8$. 
The large-$N$
treatment follows by letting $N_f \to \infty$ while keeping $\alpha$ fixed
(large but finite)~\footnote{In the physical models of \cite{dorey,fm} 
$N_f =2$, which is within the range of $N , N_c \simeq 3-4$~\cite{largen}
that allows dynamical fermion mass generation $m \ll \alpha$.}.
Notice that in this way one introduces the ratio of the 
bare coupling $g^2$ over the fixed scale $\alpha$, $g^2/\alpha \propto 1/N_f \ll 1$ as a small dimensionless expansion parameter in the analysis, and hence
one can follow the spectral representation method described above 
which is strictly 
valid for weak (dimensionful) coupling $g^2$. However, we stress
that, as compared with the dynamically generated fermion mass $m$,
the fixed scale 
$\alpha /m \gg1$, hence this is not an expansion parameter. 

To leading order in resummed one loop (leading $1/N_f$) 
analysis one obtains a 
modified statistical photon propagator:
\begin{equation}
\Delta _{\mu\nu} (k^2) = \frac{g_{\mu\nu}}{k^2\left( 1 - \Pi(k^2)\right)}
\label{photonlargen} 
\end{equation}
with $\Pi (k^2) \sim c_1\alpha /k + {\cal O}(1/N_f)$, 
to leading order in $1/N_f$ 
where $c_1$ is a numerical 
constant ($N_f$ dependent) computed in \cite{largen}. 
This implies a dressed photon propagator with softened infrared behaviour
which is formally similar to (\ref{dressphotpropapprox})
upon the substitution of $g^2$ by the fixed scale $\alpha$, however 
here the result resums one loop corrections, and hence is an 
improved version appropriate for strong coupling. 

The dressed fermion propagator is:
\begin{equation} 
S(p) = \frac{1}{A(p)}\frac{i\br p + m(p)}{p^2 -m^2(p)}
\label{fermionprop}
\end{equation}
The fermion wavefunction renormalisation $A(p)$ 
has been worked out in 
large-$N$ treatments~\cite{largen}, 
and in fact there are ambiguities as to whether it should
vanish as $p \to 0$~\cite{mavpap,largen}.
However,  as we shall discuss below, the 
explicit form of $A(p)$ will not be relevant for our purposes, 
due to 
the fact that the $A(p)$ factors  cancel out in the expression for the 
statistical-photon dressed electromagnetic vertex function, as a result
of the use of dressed vertex functions for the statistical 
photon~\cite{largen},
$\Gamma_s^\mu (p,q,k) =\frac{A(p^2) + A(q^2)}{2}\gamma^\mu$,
where p,q refer to fermion lines. 

Indeed, the electromagnetic vertex function
used for the computation of the magnetic moment in the 
model with a statistical and a real photon, reads 
in the infrared regime, where $p \simeq q$, $k \to 0$, which 
is the region responsible for the dominant contributions to 
the magnetic moment: 
\begin{equation}
\Gamma_{\rm em}^\rho = \gamma^\rho 
- i \frac{c_2}{N_f} \alpha \int \frac{d^3k}{(2\pi)^3}A(p)\gamma^\mu S(p-k)\gamma^\rho S(p-k)\gamma_\mu A(p) \frac{1}{k^2(1 - \Pi (k^2))}
\label{vertex}
\end{equation}
where $c_2$ a numerical coefficient, of ${\cal O}(1)$, 
and above we have 
explicitly  demonstrated the $1/N_f$ dependence of the 
corrections to the electromagnetic vertex, which is due to the fact 
that, in contrast to the case of 
fermion loops in the $1/N_f$-corrections to the statistical photon propagator, 
there is no summation over internal fermion lines 
at the vertex involving the real electromagnetic field. Thus the corrections 
are proportional to the  
coupling $g^2=8\alpha/N_f$ and {\it not} to the fixed scale $\alpha$. 
In the infrared limit $p-k \to p$ one can easily infer from (\ref{fermionprop})
the above-mentioned cancellation of wavefunction renormalisation 
$A(p)$ factors. Notice that this is due to the fact that 
the real electromagnetic vertex in (\ref{vertex}) is not dressed, 
in contrast with the statistical photon vertex, since $e^2 \ll \alpha$.

In the approximation where $m(p) \simeq m(0) \equiv m$ one then obtains
an expression for the vertex formally 
similar to (\ref{init}), with the 
improved photon propagator (\ref{photonlargen}), 
which is obtained from  
(\ref{dressphotpropapprox}) upon replacing $g^2/6\pi$ by $c_1\alpha$. 
Then, applying the spectral representation technique, used previously (\ref{photprop}),(\ref{rho}), with the above replacement, we arrive at the expression 
(\ref{mubfinal})
for the magnetic moment, 
with the important difference that now it is proportional to 
$\alpha/mN_f \ll 1$, as $N_f \gg 1$, where 
$m$ denotes the dynamically generated
mass~\cite{largen}. We note here that the analysis of \cite{fkmm},
using also preliminary quenched lattice studies,  have indicated a quadratic 
scaling of the dynamical fermion mass with the weak field $B$,
for strong and intermediate statistical gauge couplings $g$ in the model, 
$m/g^2 = {\rm const} + {\cal O}(\frac{e^2B^2}{g^8})$, for $ eB \ll g^4$,
but this still need to be confirmed by more complete 
lattice analysis using fully dynamical fermions. However, the analytic
results (\ref{mubfinal}) 
on a non trivial $\mu_B$ in the limit of vanishing field intensity $B
\to 0$ derived above for strong and intermediate statistical gauge couplings
are expected to hold, at least qualitatively.

As remarked earlier, 
for weakly coupled gauge theories, 
the spectral representation analysis of \cite{jackiw},
developed through one-loop Feynman graphs,  
proves sufficient 
in dealing with the infrared infinities 
of higher-order (in $g^2$) graphs, beyond one loop. This is 
due to the smallness 
of the coupling $g^2$, as compared with a typical momentum (or mass) 
scale in the problem, which allows for a consistent perturbation
expansion over $g^2/m$ to be developed, that softens the infrared
infinities of the higher-order graphs to logarithmic.

Fortunately, such a cure of higher-order infrared infinities is also true  
for our large-$N$ strongly coupled gauge theory 
studied in this subsection.  
Indeed, in this case, 
although $\alpha/m \gg 1$, where $m$ is the dynamically
generated fermion mass~\cite{appelquist,largen}, 
and hence is not a good expansion parameter, 
infrared divergent 
higher-order graphs are of order $1/N_f^2$, and hence subleading
in the limit $N_f \gg 1$.
The self-consistency of the 
leading order large-$N$ treatment~\cite{largen}, then, can be used as an 
argument supporting the qualitative conclusions drawn from 
the leading order analysis above,  
as far as the magnetic moment calculation is concerned~\footnote{In fact, studies of higher-order $1/N^2$ corrections~\cite{nash}
have shown that their effects,
at least on the dynamical mass generation, are small (of the order of 20\%), 
so that reliable qualitative information 
on the phase structure of the model can be inferred from
a leading order $1/N$ analysis.}. 
Of course, a more complete 
method in dealing with the problem is still needed for quantitatively detailed
computations to be in place~\footnote{This is  
especially because of unresolved ambiguities as far as the 
behaviour of the fermion wavefunction renormalization $A(p)$ 
in the infrared limit is concerned. 
For instance, improved Schwinger-Dyson approaches~\cite{mavpap}  
combining large-$N$ 
treatments 
with the use of pinch techniques~\cite{pinched}
show a singular 
behaviour of $A(p)$ as $p \to 0$, in contrast with 
conventional large-$N$ treatments~\cite{largen}, where $A(p)$ 
is either regular or goes to zero in this limit.},
but this falls beyond the scope of the present
work.

The anomalous magnetic moment $\mu_B$ (\ref{mubfinal})
is well-defined from the 
point of view of (infrared or ultraviolet) 
divergences, and can be computed numerically;
In a large-$N$ treatment, to leading order
in $1/N_f$~\cite{largen}, and for weak external magnetic fields, 
the dynamical fermion mass is considered as
constant, although, as we mentioned above, 
in actual situations there 
are corrections to this constant value, exhibiting a non-trivial (quadratic)
scaling with the magnetic field~\cite{fkmm}. 
From (\ref{mubfinal}) 
we then obtain for the observable magnetic moment: 
\begin{equation} 
\mu_B(N_f) = \frac{4\alpha ^2}{c N_f m^2\pi^2} \int _0^\infty 
dy \frac{1}{y^2 + (\frac{\alpha}{2m})^2}\left(J_1(y^2) + J_2(y^2)\right)
\label{mubfinalvalue}
\end{equation}
where $y=M/m$, and $J_{1,2}$ have been defined in (\ref{j12}).
In (\ref{mubfinalvalue}) 
$c$ is the real speed of light,
which in units of the fermi velocity we are working on in this paper 
is $c \sim 10^4$. As stressed in the beginning of section 2
(c.f. (\ref{rescaled})), 
this suppression factor is essential to yield the correct order 
of magnitude of the physical
magnetic moment in our system of units, which is defined as the 
response of the system to an external electromagnetic field, whose
magnitude is normalised 
with respect to the real speed of light.

The integration over $y$ 
is finite, as can be easily seen, in {\it both} the ultraviolet and
infrared regimes of $y$~\footnote{The integrals are manifestly 
infrared ($y \to 0$) finite, the potentially dangerous ultraviolet
($y \to \infty$) divergent terms coming from the ${\cal O}\left(\beta/(\sqrt{\beta} + 2)\right)$ 
and 
logarithmic terms of $J_1$ in (\ref{j12}) upon integration over $\sqrt{\beta} 
\equiv y$, weighted by terms of the form $1/(y^2 + (\alpha/2m)^2)$.
The dangerous terms are of the form:
$I_1 + I_2$, where $I_1 = 3\int_0^\infty 
dy \frac{y^2}{(y^2  + (\alpha/2m)^2)(y + 2)}$ and 
$I_2 = \int_0^\infty 
dy \frac{y^2}{2(y^2  + (\alpha/2m)^2)}{\rm ln}\left(\frac{y}{y + 2}\right)$.
To treat their ultraviolet divergent parts properly, we make use of 
the identity (\ref{identity}) in $I_2$, which yields for the divergent parts 
of $I_2$: $I_2^{({\rm div})} = \frac{1}{2}
\int_{y \to \infty} dy {\rm ln}\left( 1 - \frac{2}{2 + y}\right) \simeq 
- \int_{y \to \Lambda} dy \frac{1}{y} \sim -{\rm ln}\Lambda $, with $\Lambda$ 
an ultraviolet cutoff for (the pure number variable) $y$. 
In a similar spirit, the ultraviolet divergent parts of $I_1$
is treated properly by first changing integration variable $y \to x = y^3$,
and then approximating the denominators by going to large $x$. This
yields $I_1({\rm div}) = \int_{x \to \Lambda} dx/x = {\rm ln}\Lambda $,
using the same ultraviolet cutoff $\Lambda$ for pure numbers.
Therefore, the ultraviolet infinities cancel out 
exactly in $I_1 + I_2$, yielding
an overall finite result for $\mu_B(N_f)$, in a large-$N$ framework.},
and the result 
can be computed numerically. 
Formally, therefore, 
$\mu_B$ is of order $1/N_f$, which is small as long as $N_f \gg 1$. 

The problem is that in practice $\alpha/m \gg 1$, since $N_f \sim 2$,
and in large N treatments $m \sim \alpha e^{2+ \delta -2\pi/\sqrt{N_c/N_f - 1}}$, where $\delta$ is an arbitrary phase, not determined to leading order
in $1/N$ expansion~\cite{appelquist,largen}, 
$3 < N_c <4$, and $N_f < N_c$ for dynamical mass generation.
In order to obtain a physically consistent result
within the $1/N_f$ expansion,  
the anomalous magnetic moment
must be a small number, much smaller than 2S 
(=1 for fermions with spin S=1/2, where 2 is the gyromagnetic ratio), 
which, however, does not follow
in practice from the leading order results in $1/N_f$, as far as 
the order of $m$ is concerned~\cite{appelquist,largen,nash},
unless one selects unnaturally large values of the phase $\delta$. 
This is the main reason why one should go beyond
the large-N treatment in order to get quantitatively reliable 
results for $\mu_B$, e.g. via pinch~\cite{pinched,mavpap} 
or other non-perturbative
techniques, 
which however lies beyond our scope here.
For us, the above large-$N$ computation of $\mu_B$ 
demonstrates (formally) a finite subleading result, of order $1/N_f$ 
in the limit of $N_f \gg 1$ and 
vanishing magnetic field, which is sufficient for our purposes in this work.

\begin{figure}[t]
\centering
\includegraphics[width=10cm]{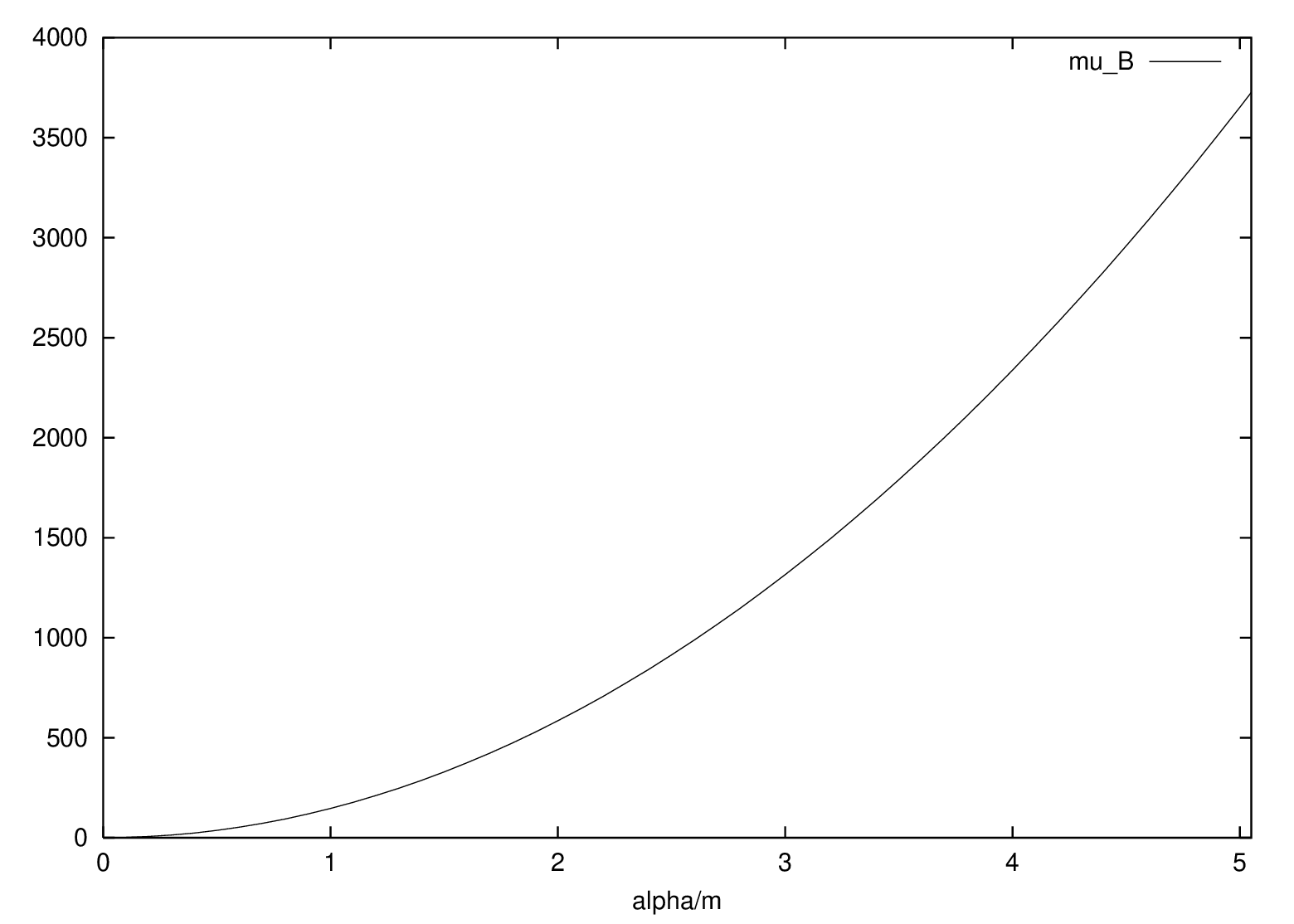}
\caption{{\it The ($c$ unscaled, for clarity in the plot) 
anomalous magnetic moment 
versus $\alpha/m$, for vanishing external field. 
The plot 
shows a quadratic dependence on $\alpha/m$. The physical 
values are obtained by the rescaling (\ref{rescaled}),
with $c \sim 10^{4}$, 
and yield reasonable, 
smaller than one, values for $\mu_B^{\rm phys}$ in the 
physical regime of fermion masses $m/\alpha \sim 0.48$. }}
\label{fig:momweak}
\end{figure}

For our purposes, therefore, we can borrow at this stage results from lattice
treatments~\cite{fkmm} of QED$_3$, for the case of $N_f = 2$, 
according to which 
$m/\alpha \sim 0.48$, which also 
agrees in order of magnitude 
with the upper bound for the dynamically generated fermion mass 
determined by the pinch technique approach~\cite{mavpap}. 
Substituting this value into (\ref{mubfinalvalue})
we then obtain an estimate of the induced anomalous magnetic moment: 
$\mu_B^{\rm phys}(N_f =2)={\cal O}(10^{-2})~<~1$ 
for the physical value $N_f = 2$. Notice that the smallness 
of the physically observable anomalous magnetic moment 
justifies the large-$N$ approximation in our model, and is due to 
the large value of the real speed of light 
$c \sim 10^4 $, which furnishes additional suppression factors
in the realistic case where the fermion flavours are $N_f=2$. 
In fig.~\ref{fig:momweak} we plot the induced anomalous magnetic moment 
(\ref{mubfinalvalue}) versus $\alpha/m$, for vanishing external field. 
The result  points towards a quadratic dependence on this parameter
$\mu_B \propto (\alpha/m)^2$, at least in the regime of 
interest in this work.

\subsection{Comparison with an Anyon Model}

Before closing this section we would like to contrast this result 
on the anomalous magnetic moment within a large-$N$ 
treatment of our parity invariant QED$_3$-like model, with a 
corresponding computation in 
an anyonic model in three-dimensions~\cite{anyon}.  
There, the authors using again a spectral representation
of the photon propagator, obtain the anomalous magnetic moment
to leading order in $1/N$ for their model, which is different 
from ours in that it consists only of a single Abelian gauge field
interacting with $2N + 1$ fermion species, in the presence
of an Abelian Chern-Simons (CS) term for the gauge field, with a coefficient
$\kappa$. In such a case, with intrinsic parity violation due to the 
dynamical CS term, the anomalous magnetic moment 
is essentially determined by the parity-violating parts of the 
dynamical gauge boson 
propagator, which are absent in our parity conserving case.
The important thing the authors of \cite{anyon} 
find is that, in the limit
where the coefficient $\kappa$ of the CS term is very large,
the induced corrections to the magnetic moment are of order $1/m\kappa$,
where $m$ is the bare fermion mass in their model. 
In fact, the result of the large-$N$ computation
of \cite{anyon} indicates a quantum corrected 
magnetic moment $\mu = \frac{1}{m}(\frac{1}{2} + \frac{1}{\kappa})$,
which, with $S = \frac{1}{2} + \frac{1}{\kappa}$ identified as the spin of
the anyon field~\cite{anyongyro}, 
implies an exact (to leading order in $1/N$) gyromagnetic ratio
$g=2$ for anyons. This result is in agreement 
with general arguments~\cite{anyongyro}
that the gyromagnetic ratio of an anyon system should remain 2 
to all orders in a quantum treatment of localised anyon fields.
Indeed, in the limit where $e^2 \kappa \gg m $,
where $e$ is the gauge coupling, 
the anyon is viewed as 
a point charge, surrounded by a gauge field cloud of size $1/e^2 \kappa$,
which collapses onto the charge, and such a system of charge-cloud composites
display a fractional statistics. 

In our case, 
things are entirely different. 
We have no anyonic fields to start with, 
hence no parity violating parts in the 
dynamical gauge boson propagator, 
and our fermions are ordinary fermions, exhibiting 
non trivial, but $1/N_f$ suppressed, corrections to their 
gyromagnetic ratio (\ref{mubfinalvalue}), responsible 
for the appearance of a parity violating term
in the effective lagrangian (\ref{mub}). This term, however, vanishes
when the external field is turned off. 
This is therefore not an intrinsic 
effect, like the one induced by the parity violating CS term
in the anyon case, but an extrinsic effect, which 
is induced by an external electromagnetic field 
that has no dynamics, and certainly no CS term.
Of course the direction of the magnetic field 
breaks time reversal, and hence parity in our 
relativistic system (due to CPT theorem), 
but this is an extrinsic breaking.
The induced anomalous magnetic moment, although  
non zero when the field is turned off, 
however, does not couple to 
the statistical gauge field so as to imply 
non trivial interactions of a permanent parity violating nature
(after the switching off of the field). The effects go away when the 
external magnetic field is switched off.
Another important point to notice is that,
unlike the anyon case,  
the sign of the induced
magnetic moment is the same for all 2$N_f$ 
two-component fermionic flavours, 
while their masses alternate sign ($N_f$ two-component 
fermions have masses $m$ and the remaining $N_f$ have masses $-m$). 
This completes our discussion on the comparison of our results with 
the anyonic case. 

Our analysis in this work deals with parity invariant QED$_3$ models with 
one type of statistical gauge coupling among the fermions. 
This has applications to one approach to high temperature
superconductivity~\cite{herbut}. 
Similar conclusions, however, are expected to hold in 
other parity invariant 
effective gauge theories of high temperature superconductors 
of the so-called 
$\tau_3$-QED type~\cite{dorey,fm},
where the abelian statistical gauge field is the U(1)
unbroken subgroup of a spin SU(2) local gauge group, and couples
to the pertinent fermion species
with opposite sign couplings, expressing the underlying 
antiferromagnetic structure of the microscopic model. 
We reserve a detailed discussion of such issues for a forthcoming 
publication. 

This completes our discussion on the weak magnetic field case. 
We next proceed to discuss the effects of a strong external field
on induced corrections to the magnetic moment in our model, which 
as we shall show scale non-trivially with the external field.

\section{Magnetic moment in strong field}

If we consider a strong external magnetic field, 
the magnetic moment is defined by computing the fermion self energy \cite{newton} (see fig. \ref{fig:vertex}(b)~\footnote{We remind the reader that 
our comments in the beginning of section 2, on the addition 
to these irreducible loop corrections of reducible graphs
expressing wavefunction renormalization of the external fermion lines
(c.g. fig.~\ref{fig:vertex}(c)),
but not contributing to the magnetic moment, remain intact in the
strong-field case 
as well.}).
For a magnetic field $B$ (and coupling $e$) in the direction perpendicular to the plane,
the fermion self energy is then connected to the anomalous magnetic moment $\mu_B$ by the relation
(apart from some phase factors, as will be seen further on)
\be\label{muBB}
\Sigma_{on-shell}=\delta m +\mu_B\frac{|eB|}{2m}\gamma^0 \left(1 + 
\cdot\cdot\cdot \right)
\ee
where $\delta m$ corresponds to the correction to the mass
and the term in parentheses denotes the projection of the 
spin on the axis of the external magnetic field~\cite{gusynin}, with 
the dots representing terms including $\gamma^1,\gamma^2$. Note that both $\mu_B$ and $\delta m$ depend on $|eB|$.
In this definition one can recognize the anomalous magnetic moment term leading to the Pauli equation in the 
non-relativistic limit (we remind the reader that, in our approach, 
its physical value is determined by the rescalings (\ref{rescale2}),
(\ref{rescaled})).
Then we can compute $\Sigma_{on-shell}$ as a loop expansion, using for the fermion propagator
the proper-time representation set up by Schwinger \cite{schwinger}.
This representation allows to take into account the 
non-perturbative interaction of the fermions with the external magnetic field. In the weak field limit, 
this interaction is perturbative and consists only in one photon propagator insertion, 
i.e. the definition of $\mu_B$ using the vertex (\ref{Gamma}) is recovered. 

In the case of 
a strong magnetic field $|eB|>>m^2$, where $m$ is the fermion mass, it is sufficient to 
consider fermions in the lowest Landau level (LLL) only. 
We will discuss later the question of definition of the anomalous magnetic moment in the LLL.

We consider the gauge $A_\mu^{ext}(x)=(0,-Bx_2/2,Bx_1/2)$. The LLL fermion 
propagator is~\cite{gusynin}:
\bea\label{LLL}
S^L(x,y)&=&\exp\{iex^\mu A_\mu^{ext}(y)\}\tilde S(x-y)\nonu
\mbox{with}~~\tilde S^L(p)&=&i\exp\left\{-\frac{p_\bot^2}{|eB|}\right\}\frac{p_0\gamma^0+m}{p_0^2-m^2}
(1-i\gamma^1\gamma^2{\rm sign}(eB)),
\eea
where $p_\bot=(p_1,p_2)$. From now on, 
unless explicitly stated, we take sign$(eB$)=1.

The one-loop fermion self energy is given by \cite{tsaialone}
\bea\label{Sigma1}
\Sigma(x,y)&=&ig^2\gamma^\mu S(x,y)\gamma^\nu D_{\mu\nu}(x,y)\nonu
&=&e^{iex^\mu A_\mu^{ext}(y)}\int\frac{d^3p}{(2\pi)^3}e^{ip(x-y)}\tilde\Sigma(p),
\eea
where $D_{\mu\nu}$ is the bare photon propagator and 
\be\label{Sigma2}
\tilde\Sigma(p)=ig^2\int\frac{d^3k}{(2\pi)^3}\gamma^\mu \tilde S(k+p)\gamma^\nu D_{\mu\nu}(k).
\ee
Note that, due to the projector $(1-i\gamma^1\gamma^2)$, the
propagator $\tilde S^L(p)$ given in Eq.(\ref{LLL}) has no inverse. For this reason, in the studies of 
dynamical symmetry breaking by a magnetic field in the LLL approximation \cite{gusynin},
one cannot use a self-consistent Dyson-Schwinger equation involving the inverse of fermion
propagator. In our case, since we are not dealing with a self-consistent equation but we are 
computing a one-loop correction, we can use the expression (\ref{Sigma1}) which involves $S$ only. From this last equation we have
\bea
\Sigma(p,q)&=&\int d^3xd^3y e^{-ipx-iqy}\Sigma(x,y)\nonu
&=&\int d^3y e^{-iy(p+q)}\tilde\Sigma(p-eA^{ext}(y)),
\eea
which is the correct expression of the fermion self energy in momentum representation. For the 
anomalous magnetic moment though, it is enough to consider the computation of $\tilde\Sigma_{on-shell}$,
since $\mu_B$ is defined as a proportionality constant which is not influenced by the phase 
gauge-dependent factor $\exp\{iexA^{ext}(y)\}$, as can be seen from Eq.(\ref{Sigma1}).

Taking into account Eqs.(\ref{LLL}), (\ref{Sigma2}), we obtain in the Feynman gauge
\bea
\tilde\Sigma(p)&=&-ig^2\int\frac{d^3k}{(2\pi)^3} \exp\left\{-\frac{(k_\bot+p_\bot)^2}{|eB|}\right\}\nonu
&&\times\frac{\gamma^\mu[(k_0+p_0)\gamma^0+m](1-i\gamma^1\gamma^2)\gamma_\mu}
{[(k_0+p_0)^2-m^2][k_0^2-k_\bot^2-M^2]},
\eea
where, as in the previous section,  we assign the mass $M$ to the photon.

This last expression can be written, when $\tilde\Sigma$ is on-shell, 
\bea\label{sigmaonshell}
\tilde\Sigma_{on-shell}&=&\frac{g^2m}{(2\pi)^3}\int d^2k_\bot\exp\left\{-\frac{k_\bot^2}{|eB|}\right\}\\
&&\times\int_0^1 dx\int_{-\infty}^{+\infty} dk_3\frac{\gamma^\mu[(1-x)\gamma^0+1](1-i\gamma^1\gamma^2)\gamma_\mu}
{\{k_3^2+x^2m^2+(1-x)(k_\bot^2+M^2)\}^2},\nonumber
\eea
where $ik_3=k_0+xp_0$ and the on-shell condition $(p_0=m,p_\bot=0)$ was used (we remind that the 
energy of the $n$th Landau level satisfies $p_0^2=m^2+2n|eB|$). This on-shell condition is not Lorentz 
invariant and this is the price to pay to define the anomalous magnetic moment in the LLL.
It has indeed been pointed out in \cite{newton,tsai,baier} that no anomalous magnetic moment in the LLL can be 
defined in a Lorentz invariant way. These authors used the equation of motion 
of the fermion for the on-shell condition, which is obviously Lorentz invariant. As a consequence, 
all reference to $\gamma^0$ disappears in the LLL and they can only define the mass shift. In our case,
this mass shift corresponds to the term which remains after taking the trace of Eq.(\ref{sigmaonshell}).

The integration over $k_3$ is straightforward and gives
\bea\label{Sigmaa}
\tilde\Sigma_{on-shell}&=&\frac{g^2m}{16\pi^2}\int d^2k_\bot\exp\left\{-\frac{k_\bot^2}{|eB|}\right\}\\
&&\times\int_0^1 dx\frac{\gamma^\mu[(1-x)\gamma^0+1](1-i\gamma^1\gamma^2)\gamma_\mu}
{\{x^2m^2+(1-x)(k_\bot^2+M^2)\}^{3/2}}.\nonumber
\eea
One can see that, in the limit $M\to 0$, a logarithmic IR divergence 
will occur. This
divergence will be treated 
in the same way as in the previous section.  
To this end consider first the naive magnetic moment
derived from the definition (\ref{muBB}). 
The quantity $\tilde\Sigma_{on-shell}$ can be written as
\begin{eqnarray}
&&\tilde\Sigma_{on-shell} =\delta m+\mu_B\frac{|eB|}{2m}\gamma^0(1+3i\gamma^1\gamma^2) \nonumber \\
&& = \delta m - \mu_B\frac{|eB|}{2m}\gamma^0\left( [1 - i\gamma^1\gamma^2
{\rm sign}(eB)] - 2[1 + i\gamma^1\gamma^2{\rm sign}(eB)]\right) 
\end{eqnarray}
Note that in the last line we have re-written the right hand side 
in such a way so as to
make clear that, when one calculates {\it on-shell} quantities 
entering the expression for the anomalous magnetic moment
(see below), terms proportional to $1 + i\gamma_1\gamma_2{\rm sign}(eB)$ do not contribute, because in the LLL approximation 
the spin is polarised along the magnetic field~\cite{gusynin}.

The anomalous magnetic moment $\mu_B$ is then 
\bea
\mu_B^{(0)}(m,M)&=&\frac{g^2m^2}{8\pi}\int_0^\infty du~e^{-u}\nonu
&\times&\int_0^1 dx\frac{1-x}{\{x^2m^2+(1-x)(u|eB|+M^2)\}^{3/2}}.
\eea

\vspace{.5cm}

We first note that in the case $m=0$ and $M\ne 0$, the integration over $x$ converges and the 
final result for the magnetic moment gives $\mu_B^{(0)}(0,M)=0$. 

In the other situation, where $m\ne 0$ and $M\to 0$, we first consider $M\ne 0$ and write 
\be
\mu_B^{(0)}(m,M)=\frac{g^2}{8\pi m}\int_0^\infty due^{-u}\int_0^1 dx
\frac{1-x}{\{x^2+(1-x)(\alpha u+\beta)\}^{3/2}},
\ee
where $\alpha=|eB|/m^2>>1$ and $\beta=M^2/m^2$.

We first perform the integration over $u$, using the fact that $\alpha>>1$:
we can approximate 
\bea
&&\int_0^\infty due^{-u}\frac{1-x}{\{x^2+(1-x)(\alpha u+\beta)\}^{3/2}}\nonu
&=&\int_0^1du\frac{1-x}{\{x^2+(1-x)(\alpha u+\beta)\}^{3/2}}+{\cal O}(1/\alpha)^{3/2}\nonu
&=&\frac{2/\alpha}{\sqrt{x^2+(1-x)\beta}}+{\cal O}(1/\alpha)^{3/2}.
\eea  
The remaining integral over the Feynman parameter $x$ gives then for any non-vanishing $\beta$
\be\label{massphot2}
\mu_B^{(0)}(m,M)=\frac{g^2m}{4\pi|eB|}\ln\left(\frac{2m+M}{M}\right)+{\cal O}(m/\sqrt{eB})^3.
\ee

\vspace{.5cm}

Finally, the divergence-free anomalous magnetic moment $\mu_B(m,0)$ is obtained with the same method
as in the previous section:
\be\label{magmomint}
\mu_B(m,0)=\int_0^\infty dM\rho(M)\mu_B^{(0)}(m,M),
\ee
where $\rho(M)$ is given by Eq.(\ref{rho}). We neglect here the influence of the magnetic field
on the fermion loop, which corresponds to higher order contributions.

For the integration over $M$, we consider two cases
which are relevant:
\begin{itemize}
\item in the limit $g^2<<m$:
\be\label{muBstrongB1}
\mu_B^{(1)}(m,0)=\frac{g^2m}{4\pi|eB|}\ln\left(\frac{12\pi m}{g^2}\right)+{\cal O}(g^2/m);
\ee
\item in the limit $m<<g^2$:
\be\label{muBstrongB2}
\mu_B^{(2)}(m,0)=\frac{6m^2}{\pi|eB|}\left\{1-\ln\left(\frac{12\pi m}{g^2}\right)\right\}+{\cal O}(m/g^2).
\ee
\end{itemize}

Note that among the three mass scales $(\sqrt{|eB|},g^2,m$), 
$\sqrt{|eB|}$ is the largest one, such that $\mu_B$ is always perturbative.
Once again, we remind the reader that 
in order to give the physically correct order of magnitude,
at least for potential applications to the theory of
high temperature superconductivity, 
we should 
use the rescaling (\ref{rescale2}),(\ref{rescaled}).

We shall conclude this section by discussing 
the scaling of the anomalous magnetic moment $\mu_B$ with the 
externally applied magnetic field $B$. 
It was found in \cite{fkm} that the fermion mass generated dynamically in the presence of a 
strong magnetic field is 
\be\label{mdyn}
m_{dyn}\simeq Cg^2\ln\left(\frac{\sqrt{|eB|}}{g^2}\right),
\ee
where $C$ is a constant. This 
dependence has been seen numerically \cite{afk}, where it was found that $C\simeq 0.06$.

\begin{figure}[t]
\centering
\includegraphics[width=8cm]{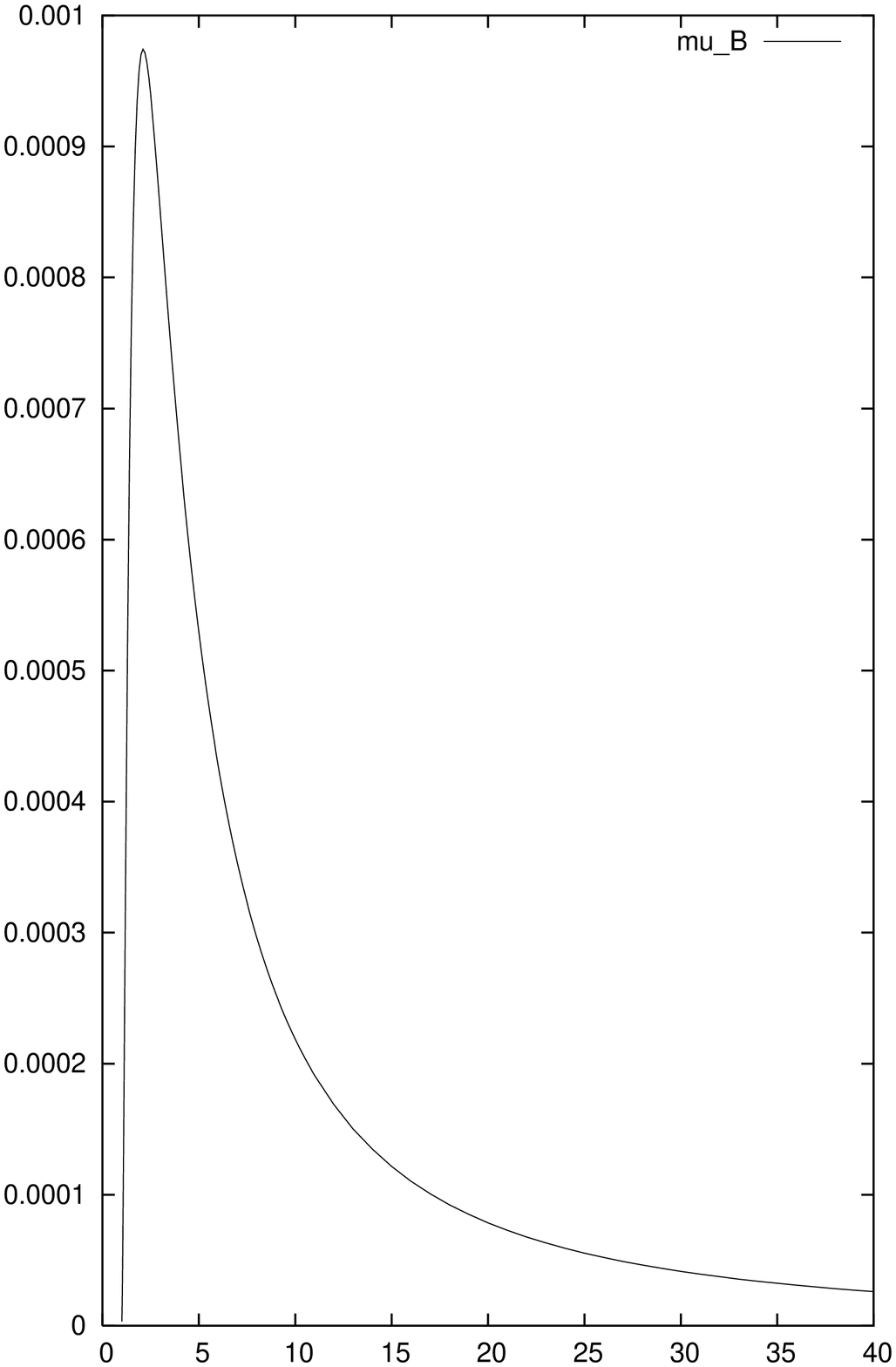}
\caption{{\it The anomalous magnetic moment versus the dimensionless field strength $b$ 
shows a maximum for $b\simeq 2$.}}
\label{fig:magmom}
\end{figure}

Let us define the dimensionless magnetic strength $b=\sqrt{|eB|}/g^2$.
The lowest Landau level approximation assumes that $m_{dyn}<<\sqrt{|eB|}$, which from Eq.(\ref{mdyn})
and in terms of $b$ reads 
\be
C\ln b<<b.
\ee
This is always the case, independently of the strength of the coupling $g$.  
Making the identification $m=m_{dyn}$ in the integral (\ref{magmomint}), we obtain the following 
expression for $b$-dependence of the anomalous magnetic moment
(using the rescaling (\ref{rescale2}),(\ref{rescaled}), in order
to give the physically correct order of magnitude):
\be
\mu_B(b)=\frac{1}{c}\frac{C}{2\pi^2}\frac{\ln b}{b^2}\int_0^\infty dx\frac{\ln(x+\kappa\ln b)}{1+x^2},
\ee
where the constant is $\kappa=12\pi C\simeq 2.26$.
We plot the $c$-unscaled anomalous magnetic moment 
$c\mu_B(b)$ versus $b$ in 
fig.~\ref{fig:magmom}. We observe  that the anomalous magnetic 
moment first increases very quickly for $1<b<2$ to reach a maximum 
when $b\simeq 2$, and then decreases 
as $1/b^2$ for $b>2$.

This behaviour can be tested experimentally, 
in case we apply such theoretical models to the physics of high temperature 
planar superconductors or, more generally, planar doped antiferromagnets. 
We reserve a detailed analysis of such physical applications
for a future publication. However, for completeness we would like to 
give at this stage 
an estimate of the magnetic field for which the maximum of the 
anomalous magnetic moment occurs in the QED-like model for high temperature
superconductivity of \cite{fm}. In those works we have a relativistic 
gauge system of nodal holon fermions coupled to a statistical gauge field,
with a nodal velocity (playing the role of the speed of light) 
$\hbar v_F = 5 \cdot 10^{-4} c$, where $c$ is the real speed of light. 
The statistical gauge coupling $g^2$ is estimated to be~\cite{fm} 
$g^2 \sim 4\pi \eta J \hbar v_F $, where $\eta$ is the doping 
concentration in the sample, and $J$ is the Heisenberg 
magnetic interaction. For maximum doping concentrations of order 10\%,
we have~\cite{fm}: $\eta J \sim {\cal O}(10)~$meV, from which 
we obtain the estimate:
\begin{equation} 
g^2 \sim 6.3 \cdot 10^{-15}~{\rm GeV}
\label{gestimate}
\end{equation} 
Thus, the maximum of $\mu_B$, which, as we infer 
from fig.~\ref{fig:magmom}, occurs 
for $b=\sqrt{eB}/g^2 \simeq 2$,
is attained for magnetic fields of order $B \sim 6 \cdot 10^{-9}~$Tesla =
0.06~mGauss. These are still large compared to the order of the 
magnetically induced mass gaps (\ref{mdyn}) 
of fermion excitations at the nodes of this specific model~\cite{fm}.
Note that in condensed matter situations one uses external fields
usually from the mGauss range 
up to 10~Tesla~\cite{fm}. 

Experimentally, therefore, 
in order to test the predictions of this analysis on $\mu_B$, 
should the model be realised in nature,
one should look at excitations near the nodes
of the superconducting gap, 
within the superconducting (holon gapped) phase
of a high-temperature superconducting planar material,  
and 
vary
the
externally applied magnetic field around the very low range 
of $0.06$ mGauss. According to the model presented here, then,
one should   
observe (c.f. fig.~\ref{fig:magmom})
a sharp increase (by an order of magnitude) and eventual decrease
of the induced anomalous magnetic moment of the nodal 
holon degrees of freedom. 
For larger magnetic fields, of the order often used in condensed
matter applications, that is up to a few Tesla,  
the induced $\mu_B$ is considerably smaller, and there is a smooth
dependence on the field intensity, as shown in fig.~\ref{fig:magmom}. 
We intend to come back to a 
detailed analysis of such issues, in the context of realistic 
condensed matter models, in a future publication.

\section{Conclusions and Outlook} 

In this work we have discussed the anomalous magnetic moment of 
fermions in parity conserving QED$_3$, induced by the 
application of external fields. The induced term scales non trivially
with the magnetic field, and when the latter is turned off to zero,
one is left with a non trivial residual magnetic moment (\ref{muB}).
On the other hand, in the presence of a strong field the induced magnetic moment scales non 
trivially with the field's intensity (\ref{muBstrongB1}), (\ref{muBstrongB2}).

These results may be of relevance to the physics of high-temperature
superconducting materials, where, upon the application of a 
strong external magnetic field one 
observes indication for parity violation.  
In our model above we did not specify the physical nature of the 
dynamical ``photon'', which thus could be the statistical gauge field
of \cite{dorey,fm,herbut} representing magnetic interactions (pairing) causing 
superconductivity. In fact we have been careful to keep its 
(three-dimensional) coupling $g^2$ 
different from the electromagnetic charge $e$ appearing in front of 
terms pertaining to the external field. 

The induced parity violation is due to the magnetic moment term and not due 
to the mass gap of the fermions, which remains parity invariant. 
The derived scaling with the magnetic field can be tested experimentally.
Moreover, the way we derived such a scaling, by means of 
inserting a mass for the dynamical ``photon'' in order to 
regularize IR divergences acquires itself a physical 
significance, once we apply the model to the physics of planar 
superconductors. Indeed, in 
models such as \cite{dorey,fm}, the statistical ``photon'', 
which in our discussion above appears with a coupling $g^2$, 
may become massive (with a non-perturbatively small mass) 
in the pseudogap phase, due to non perturbative
effects~\cite{asm}, while it is exactly massless in the superconducting phase.
The different scaling of the induced magnetic moment of the holons then
between the two cases (c.f. [(\ref{massphot1}), (\ref{massphot2})] and 
[(\ref{mubfinalvalue}),(\ref{muBstrongB1}),(\ref{muBstrongB2})]), 
discussed in this work, may provide a
way of testing such models experimentally. 

There are many issues that we should look at in the near future.
First of all, one should use realistic effective theories of high temperature
superconductors, including the effects of holons. In a recent work~\cite{asm}
we have conjectured that there are regions in the parameter space of 
the condensed-matter models where spinons and holons exhibit (extended) 
supersymmetric
dynamics. The physical degrees of freedom in such models are composites
of spinons and holons, and they themselves exhibit supersymmetric 
spectra. In such cases one has both bosonic and fermionic charged excitations,
which can both couple to the magnetic field. However the presence of 
the field breaks supersymmetry explicitly, due to the different
ways by which bosons and fermions couple to a magnetic field. 
It would be interesting 
to study the induced magnetic moment in such broken supersymmetric cases,
especially from the point of view of scaling with the external field.
In fact, in four-dimensional particle physics, such computations 
reveal important signatures of supersymmetry, which can distinguish 
the Standard Model from, say, the Minimal Supersymmetric one. 
In a similar spirit, we expect in three dimensions important 
differences between the supersymmetric and non supersymmetric 
composite field theories describing the effective dynamics of
(some regions of the parameter 
space of) antiferromagnets, which could be accessible to experiment.
We plan to discuss such issues in the near future. 

\section*{Acknowledgements} 

The authors wish to thank A. Kovner and J. Polonyi  
for detailed discussions on the spectral representation method. 
K.F. and N.E.M. wish to thank CERN-Physics-Theory 
and the Department of Theoretical Physics of 
the University of Valencia (Spain), respectively, for the hospitality
during the last stages of this work.

\end{document}